\documentclass[twocolumn,showpacs,superscriptaddress, preprintnumbers , amsmath,amssymb,letterpaper]{revtex4}

\usepackage{dcolumn}
\usepackage{bm}
\usepackage{amsmath}
\usepackage{multirow}
\usepackage{graphicx}
\usepackage{float}
\usepackage[colorlinks=true]{hyperref}

\begin{document}


 \newcommand{\re}{\mathop{\mathrm{Re}}}
 \newcommand{\im}{\mathop{\mathrm{Im}}}
 \newcommand{\D}{\mathop{\mathrm{d}}}
 \newcommand{\I}{\mathop{\mathrm{i}}}
 \newcommand{\E}{\mathop{\mathrm{e}}}
 \newcommand{\unite}[2]{\mbox{$#1\,{\rm #2}$}}
 \newcommand{\myvec}[1]{\mbox{$\overrightarrow{#1}$}}
 \newcommand{\mynor}[1]{\mbox{$\widehat{#1}$}}
 \newcommand{\rmsemit}{\mbox{$\tilde{\varepsilon}$}}
 \newcommand{\mean}[1]{\mbox{$\langle{#1}\rangle$}}

\title{Generation and Characterization of Electron Bunches with Ramped \\
 Current Profiles in a Dual-Frequency Superconducting Linear Accelerator}
\author{P. Piot} \affiliation{Northern Illinois Center for
Accelerator \& Detector Development and Department of Physics,
Northern Illinois University, DeKalb IL 60115,
USA} \affiliation{Accelerator Physics Center, Fermi National
Accelerator Laboratory, Batavia, IL 60510, USA}
\author{C. Behrens} \affiliation{Deutsches Elektronen-Synchrotron DESY,  Notkestra\ss{e} 85
D-22607 Hamburg, Germany}
\author{C. Gerth} \affiliation{Deutsches Elektronen-Synchrotron DESY,  Notkestra\ss{e} 85
D-22607 Hamburg, Germany}
\author{M. Dohlus} \affiliation{Deutsches Elektronen-Synchrotron DESY,  Notkestra\ss{e} 85
D-22607 Hamburg, Germany}
\author{F. Lemery} \affiliation{Northern Illinois Center for
Accelerator \& Detector Development and Department of Physics,
Northern Illinois University, DeKalb IL 60115, USA}
\author{D. Mihalcea}\affiliation{Northern Illinois Center for
Accelerator \& Detector Development and Department of Physics,
Northern Illinois University, DeKalb IL 60115, USA}
\author{P. Stoltz}\affiliation{Tech-X Corporation, Boulder, CO 80303, USA}
\author{M. Vogt} \affiliation{Deutsches Elektronen-Synchrotron DESY,  Notkestra\ss{e} 85
D-22607 Hamburg, Germany}
%

\preprint{DESY TESLA-FEL 11-02 and FERMILAB-PUB 11-339-APC}
\date{\today}

\begin{abstract}
We report on the successful experimental generation of electron bunches with ramped current profiles. The technique relies on impressing nonlinear correlations in the longitudinal phase space using a superconducing radiofrequency linear accelerator operating at two frequencies and a current-enhancing dispersive section. The produced $\sim 700$-MeV bunches have peak currents of the order of a kilo-Amp\`ere. Data taken for various accelerator settings demonstrate the  versatility of the method and in particular its ability to  produce current profiles that have a quasi-linear dependency on the longitudinal (temporal) coordinate.  The measured bunch parameters are shown, via numerical simulations, to produce gigavolt-per-meter peak accelerating electric fields with transformer ratios larger than 2 in dielectric-lined waveguides. 
\end{abstract}
\maketitle
Electron acceleration is a rapidly-advancing field of scientific research with widespread applications in industry and medicine. Producing and accelerating high-quality electron bunches within very compact footprints is a challenging task that will most probably use advanced acceleration methods. These techniques can be categorized into laser-driven~\cite{malka,leemann,varin} and charged-particle-beam-driven methods~\cite{pisin,petra,gai,Caldwell}. In the latter scheme, a popular configuration consists of a ``drive" electron bunch with suitable parameters propagating through a high-impedance structure or plasma medium thereby inducing an electromagnetic wake. A following ``witness" electron bunch, properly delayed, can be accelerated by these wakefields. 

Collinear beam-driven acceleration techniques have demonstrated accelerating fields in excess of GV/m~\cite{slac,matt}. The fundamental wakefield theorem~\cite{wake} limits the transformer ratio -- the maximum accelerating wakefield over the decelerating field experienced by  the driving bunch -- to 2 for bunches with symmetric current profiles. Tailored  bunches with  asymmetric , e.g. a linearly-ramped,  current profiles can lead to transformer ratio $> 2$~\cite{bane}. 
To date, there has been a small number of techniques capable of producing linearly-ramped electron bunches.  A successful experiment demonstrated the production of 50-A ramped electron bunches using sextupole magnets located in a dispersive section~\cite{england} to impart nonlinear correlation in the longitudinal phase space (LPS).  Unfortunately, the method  introduces coupling between the longitudinal and transverse degrees of freedom which ultimately affects the transverse brightness of the drive and witness bunches. 

In this Letter we present an alternative technique that uses a radiofrequency (rf) linear accelerator (linac) operating at two frequencies.  It has long been recognized that linacs operating at multiple frequencies could be used to correct for LPS distortions and improve the final peak current~\cite{smith,dowell}.  We show analytically and demonstrate experimentally how a two frequency linac could be operated to tailor the nonlinear correlations in the LPS thereby providing  control over the current profile. 

\begin{figure}[hhhhh!!!!!!!!!!!!]
\centering
\includegraphics[width=0.46\textwidth]{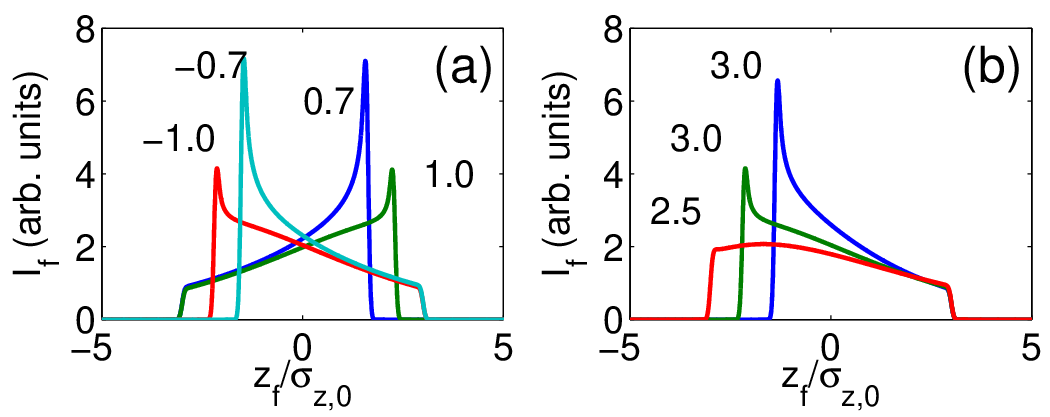}
\caption{(color online) Analytically-computed current profiles for several values of $b_f$ for fixed $a_f=2.5$ (a) and for several values of $a_f$ with $b_f=0.7$ (b). The numbers in (a) [resp. (b)] are the values of $b_f$ [resp. $a_f$]; for all the cases  $\sigma_u=0.05$. }
\label{fig:current}
\end{figure}

We first elaborate the proposed method using a 1D-1V single-particle model of the LPS dynamics and take an electron with coordinates $(z,\delta)$ where $z$ refers to the longitudinal position of the electron with respect to the bunch barycenter (in our convention $z>0$ corresponds to the head of the bunch) and $\delta\equiv p/\mean{p}-1$ is the fractional momentum spread ($p$ is the electron's momentum and $\mean{p}$ the average momentum of the bunch).   Considering a photo-emission electron source, the LPS coordinates downstream are $(z_0, \delta_0=a_0 z_0 + b_0 z_0^2 + {\cal O}(z_0^3))$ where $a_0$, and $b_0$ are constants that depend on the bunch charge and operating parameters of the electron source. For sake of simplicity we limit our model to second order in $z_0$ and $\delta_0$.  Next, we examine the acceleration through a linac operating at the frequencies $f_1$ and $f_n\equiv n f_1$ with total accelerating voltage $V(z)=V_1\cos(k_1z+\varphi_1) + V_n \cos(k_n z+\varphi_n) $ where $V_{1,n}$ and  $\varphi_{1,n}$ are respectively the accelerating voltages and operating phases of the two linac sections, and $k_{1,n}\equiv2\pi f_{1,n}/c$. In our convention, when the phases between the linac sections and the electron bunch are $\varphi_{1,n}=0$ the bunch energy gain is maximum (this is refer to as on-crest operation). Under the assumption $k_{1,n}z_0 \ll 1$ and neglecting non-relativistic effects, the electron's LPS coordinate downstream of the linac are $(z_l=z_0, \delta_l=a_l z_0+b_lz_0^2)$ where  $a_l\equiv a_0-e(k_1 V_1 \sin\varphi_1 + k_n V_n \sin\varphi_n )/\bar{E}_l$, $b_l\equiv b_0-e(k_1^2 V_1 \cos\varphi_1 + k_n^2 V_n \cos\varphi_n )/(2 \bar{E}_l)$ with $e$ being the electronic charge and  $\bar{E}_l$ the beam's average energy downstream of the linac. Finally, we study the passage of the bunch through an achromatic current-enhancing dispersive section [henceforth referred to as ``bunch compressor" (BC)]. The LPS dynamics through a BC is approximated by the transformation $z_f= R_{56}\delta_l +T_{566} \delta_l^2 $ where $R_{56}$ (also referred to as longitudinal dispersion), and $T_{566}$ are the coefficients of the Taylor expansion of the transfer map $\langle z_f | \delta_l \rangle$ of the BC. 
Therefore the final position is given as function of the initial coordinates following $z_f = a_f z_0 + b_f z_0^2 $ with $a_f \equiv 1 + a_l R_{56}$ and $b_f\equiv b_l R_{56}+ a_l^2 T_{566}$. Taking the initial current  to follow the  Gaussian distribution  $I_0(z_0)=\hat{I_0} \exp [-z_0^2/(2\sigma_{z,0}^2)]$ (where $\hat{I_0}$ is the initial peak current), and invoking the charge conservation $I_f(z_f) dz = I_0(z_0) dz_0$ gives the final current distribution $I^u_f(z_f)=\hat{I}_0/\Delta^{1/2}(z_f) \exp[-(a_f + \Delta^{1/2}(z_f))^2/(8b_f^2 \sigma_{z,0}^2)]  \Theta [\Delta(z_f)]$ where $\Delta(z_f) \equiv a_f^2+4b_f z_f $ and $\Theta()$ is the Heaviside function. The latter current distribution does not include the effect of the initial uncorrelated fractional momentum  spread $ \sigma^u_{\delta,0}$.  The final current, taking into account  $ \sigma^u_{\delta,0}$, is given by  the convolution $I_f(z_f)=\int d\tilde{z_f} I^u_f(\tilde{z_f}) \exp[-(z_f-\tilde{z_f})^2/(2\sigma_u^2)]$ where $\sigma_u \equiv  R_{56} \sigma^u_{\delta,0}$. The final current shape is controlled via  $a_f$ and $b_f$ and can be tailored to follow a  linear ramp as demonstrated in Fig.~\ref{fig:current}.

\begin{figure}[hh!!]
\includegraphics[width=0.49\textwidth]{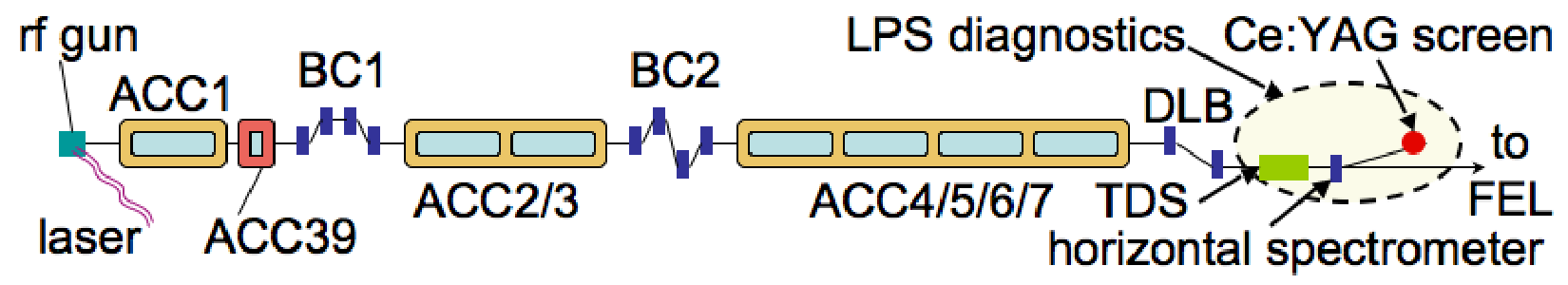}
\caption{(color online) Diagram of the FLASH facility. Only components affecting the longitudinal phase space beam (LPS) dynamics of the bunches are shown. The acronyms ACC, BC, and DBL stand respectively for accelerating modules,  bunch compressors, and dogleg beamline (the blue rectangles represent dipole magnets). The transverse deflecting structure (TDS), spectrometer and Ce:YAG screen compose the LPS  diagnostics. }
\label{fig:flash}
\end{figure}

The experiment described in this Letter was performed at the Free-electron LASer in Hamburg (FLASH) facility~\cite{flash}. In the  FLASH accelerator, diagrammed in Fig.~\ref{fig:flash}, the electron bunches are generated via photoemission from a cesium telluride photocathode located on the back plate of a 1+1/2 cell normal-conducting rf cavity operating at 1.3 GHz on the TM$_{010}$ $\pi$-mode (rf gun).  The bunch is then accelerated in a 1.3-GHz and 3.9-GHz superconducting accelerating modules (respectively ACC1 and ACC39) before passing through a bunch compressor (BC1). The ACC39 3rd-harmonic module was installed to nominally correct for nonlinear distortions in the LPS and enhance the final peak current of the electron bunch~\cite{piot}. Downstream of BC1, the bunch is accelerated and can be further  compressed in BC2. A last acceleration stage (ACC4/5/6/7) brings the beam to its final energy (maximum of $\sim 1.2$~GeV). The beam's direction is then  horizontally translated using a dispersionless section referred to as dogleg beamline (DLB). Nominally, the beam is sent to a string of undulators to produce ultraviolet light via the self-amplified stimulated emission free-electron laser (FEL) process. For our experiment, the bunches were instead vertically sheared by a 2.856-GHz transverse deflecting structure (TDS) operating on the TM$_{110}$-like mode and horizontally bent by a downstream spectrometer~\cite{beherens}. Consequently the transverse density measured on the downstream Cerium-doped Yttrium Aluminum Garnet (Ce:YAG) scintillating screen is representative of the LPS density distribution. The horizontal and vertical coordinates at the Ce:YAG screen are respectively  $x_s \simeq \eta \delta_F$, where $\eta\simeq 0.75$~m is the horizontal dispersion function, and $y_s \simeq \kappa z_F$ where $\kappa \simeq 20$ is the vertical shearing factor and $(z_F, \delta_F)$ refers to the LPS coordinate upstream of the TDS.  The exact values of $\eta$ and $\kappa$ are experimentally determined via a beam-based calibration procedure. 

\begin{table}[hbt]
\caption{Settings of accelerator subsystems relevant to the LPS dynamics used in the experiment and simulations.\label{tab:settings} }

\begin{center}
\begin{tabular}{l c c c}\hline\hline\
parameter & symbol & value & unit \\
\hline
ACC1 voltage & $V_1$  & [140-157] & MV  \\
ACC1 phase  &  $\varphi_1$    & [-10,10]  & deg  \\
ACC39 voltage & $V_3$  & [13,21]  & MV  \\
ACC39 phase  &  $\varphi_3$    & [160-180]  & deg  \\
ACC2/3 voltage & $V_{1,2-3}$  & 311 & MV  \\
ACC2/3 phase  &  $\varphi_{1,2-3}$    & 0  & deg  \\
ACC4/5/6/7 voltage & $V_{1,4-7}$  & 233.9 & MV  \\
ACC4/5/6/7 phase  &  $\varphi_{1,4-7}$    & 0  & deg  \\
BC1 longitudinal dispersion & $R_{56}^{(1)} $ &   $\sim 170$ & mm\\
BC2 longitudinal dispersion & $R_{56}^{(2)} $  &  $\sim 15$ & mm\\
Single-bunch charge & $Q$ & 0.5 & nC \\
Bunch energy & $E$ & $\sim 690$ & MeV \\
\hline \hline
\end{tabular}
\end{center}
\end{table}

The accelerator parameters settings are gathered in Tab.~\ref{tab:settings}. The nominal settings of BC2 were altered to reduce its longitudinal dispersion $R_{56}^{(2)}$ and the ACC2/3 and ACC4/5/6/7 accelerating modules were operated on crest. Such settings insure that the BC2  and the DBL sections do not significantly affect the LPS beam dynamics. Therefore the measured current profile is representative of the profile downstream of BC1. 

\begin{figure}[hhhhh!!!!!!!!!!!!]
\centering
\includegraphics[width=0.43\textwidth]{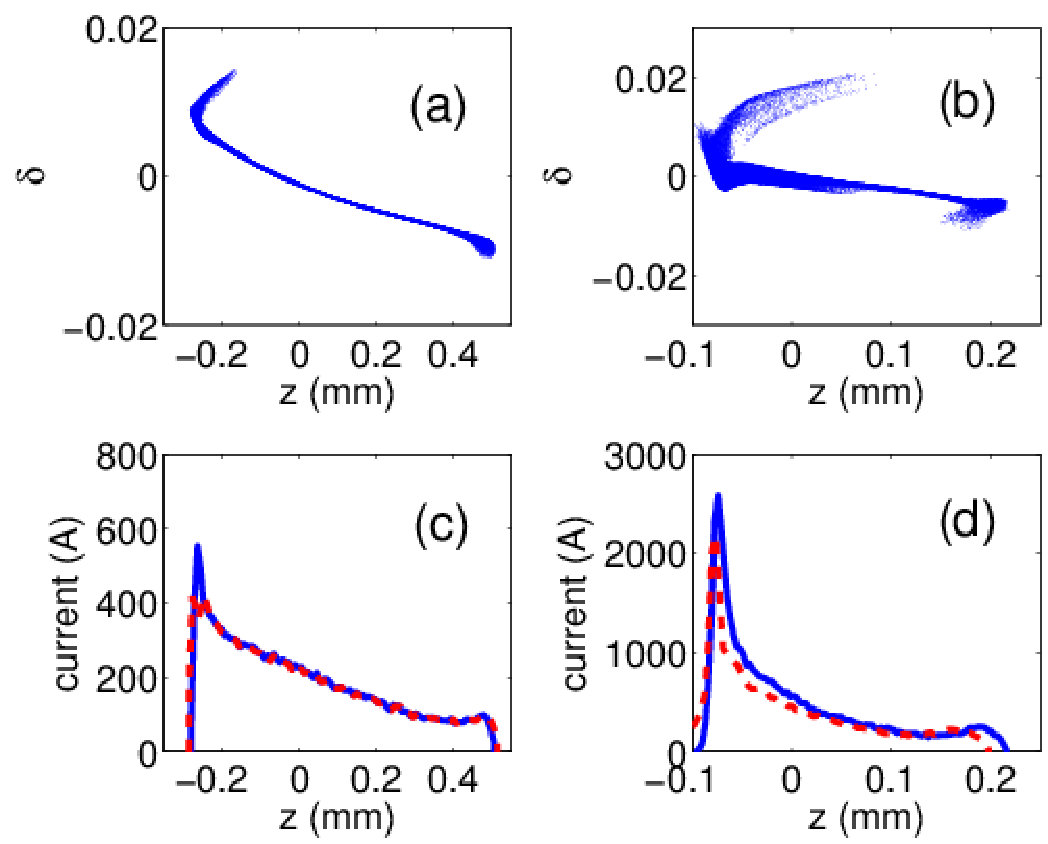}
\caption{(color online) Simulated LPS distribution [(a) and (b)] with associated current profile downstream of BC1 (solid blue trace) and DBL (dash red trace) [(c) and (d)]. The set of plots [(a), (c)] and [(b), (d)] correspond to different $(V_{1,3}, \varphi_{1,3})$ settings.  \label{fig:1d1v}}
\end{figure}

In order to validate the simple analytical model described above,  numerical simulations of the LPS beam dynamics were carried using a multi-particle model. The simulations also enable the investigation of possible detrimental effects resulting from collective effects such as longitudinal space charge (LSC) and beam self interaction via coherent synchrotron radiation (CSR)~\cite{saldin}. In these simulations, the beam dynamics in the rf-gun was modeled with the particle-in-cell program {\sc astra}~\cite{astra} and the obtained distribution was subsequently tracked in the accelerating modules using a 1D-1V program that incorporates a one-dimensional model of the LSC. The program {\sc csrtrack}~\cite{csrtrack}, which self-consistently simulates CSR effects, was used to model the beam dynamics in the BC1, and BC2 sections. An example of simulated LPS distributions and associated current profiles computed for different settings of ACC1 and ACC39 parameters appear in Fig.~\ref{fig:1d1v}. The results indicate that the production ramped bunches is possible despite the intricate LPS structures developing due to the collective effects and higher-order nonlinear effects not included in our analytical model. The simulations also confirm that the current profile upstream of the TDS (as measured by the LPS diagnostics) is representative of the one downstream of BC1.  \\

\begin{figure}[tt]
\centering
\includegraphics[width=0.435\textwidth]{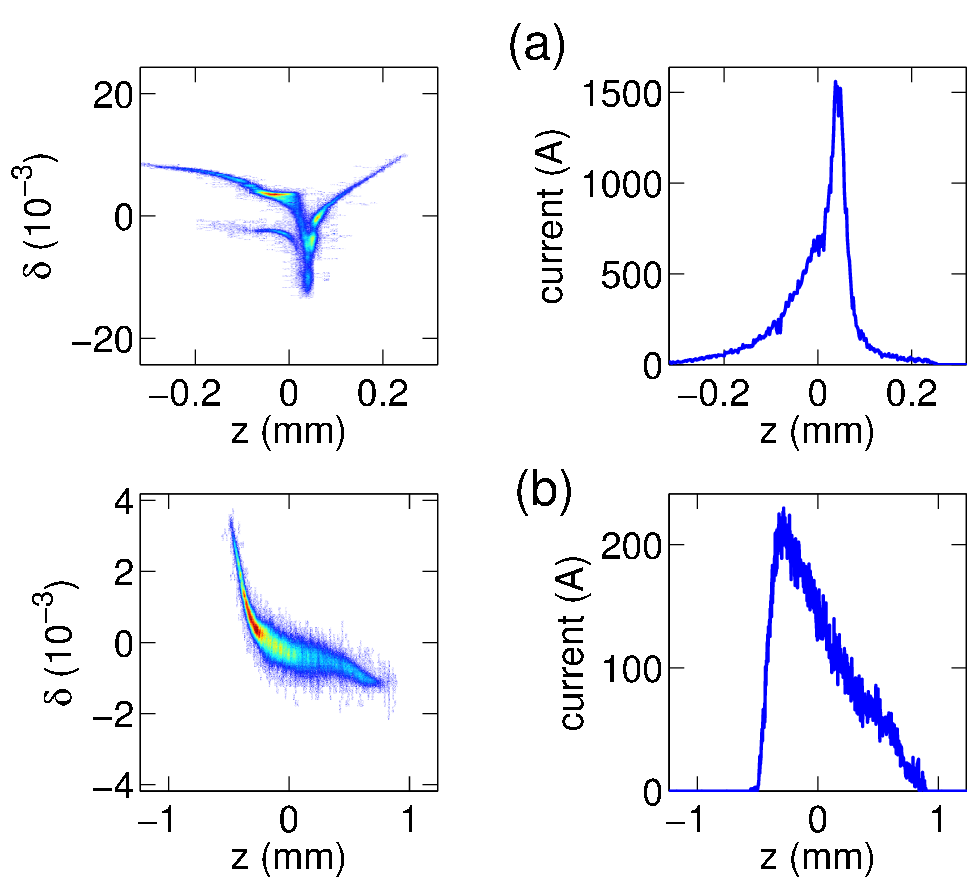}\\
~~\includegraphics[width=0.435\textwidth]{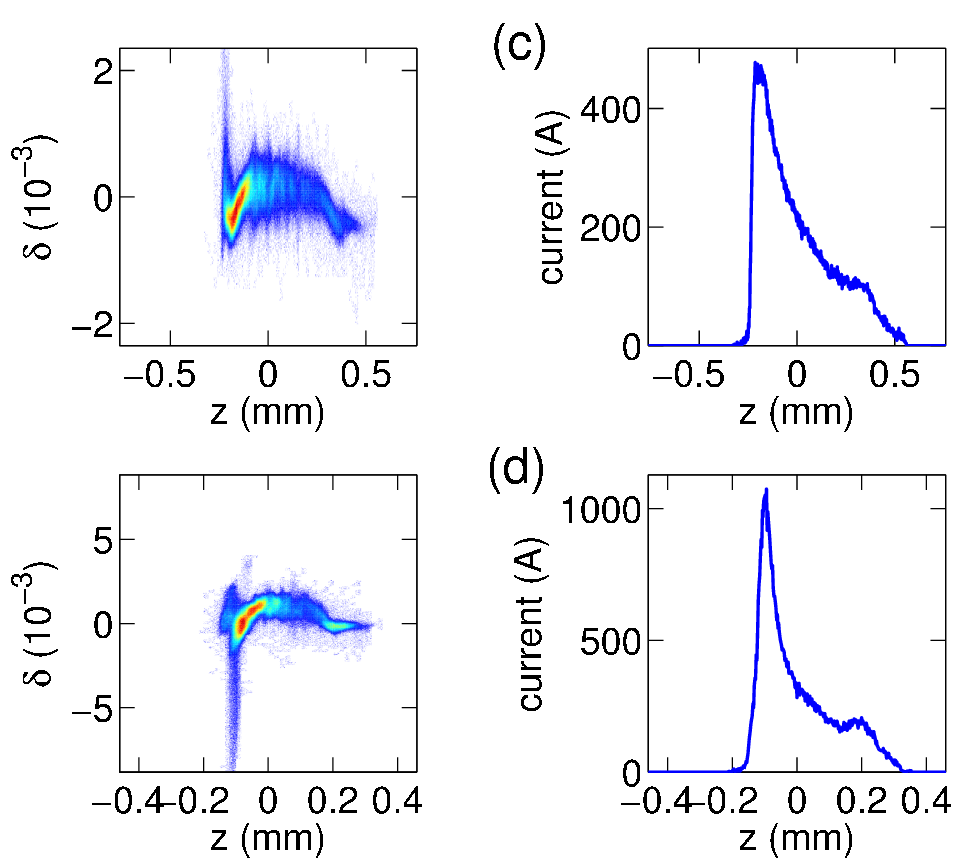}
\caption{(color online) Snapshots of the measured longitudinal phase spaces (left column) and associated current profiles (right column) for different settings of the ACC1 and ACC39 accelerating modules. The values $(V_1, \varphi_1; V_3,\varphi_3)$  [in (MV,$^{\circ}$,MV,$^{\circ}$)] are: (150.5, 6.1; 20.7, 3.8), (156.7, 3.8; 20.8, 168.2), (155.6, 3.6; 20.6, 166.7), and (156.8, 4.3; 20.7, 167.7) for respectively case (a), (b), (c), and (d).} \label{fig:exp}
\end{figure}

Figure~\ref{fig:exp}  displays examples of measured LPS distributions with associated current profiles obtained for different settings of ACC1 and ACC39.  As predicted, the observed current profiles are asymmetric and can be tailored to be ramped with the head of the bunch ($z>0$) having less charge than the tail; see Fig.~\ref{fig:exp} (b-d). The latter feature is in contrast with the nominal compression case at FLASH  where the LPS distortion usually results in a low-charge trailing population as seen in Fig.~\ref{fig:exp} (a). 

\begin{figure}[hhhhh!!!!!!!!!!!!]
\centering
\includegraphics[width=0.48\textwidth]{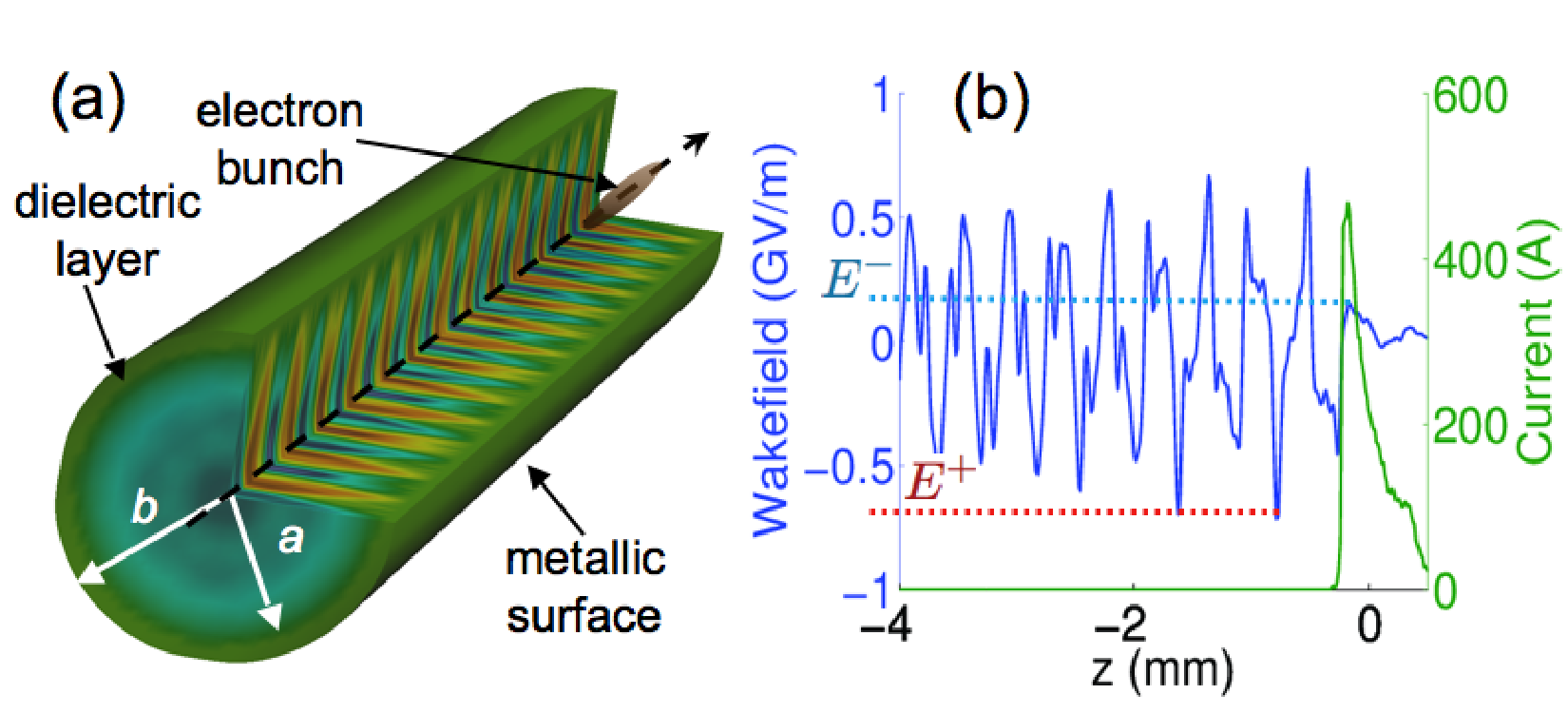} 
\caption{(color online) Cylindrical-symmetric dielectric-loaded waveguide considered (a) and  axial wakefield produced by the current profile shown in Fig.~\ref{fig:exp} (c) for ($a$, $b$)=(20,60)~$\mu$m. \label{fig:wake}}
\end{figure}
\begin{figure}[hhhh!!!!!]
\centering
\includegraphics[width=0.48\textwidth]{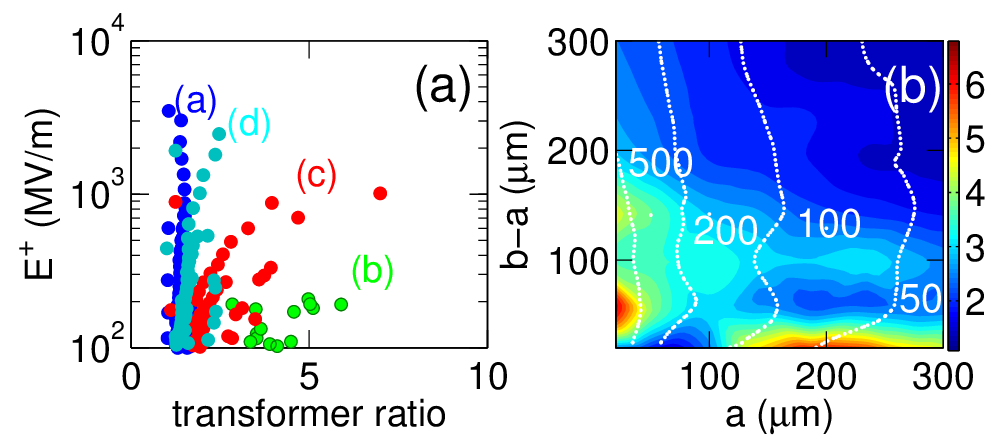} 
\caption{(color online) Simulated transformer ratio versus peak accelerating field (a) for the four measured current profiles (displayed as different colors with label corresponding to cases  shown in Fig.~\ref{fig:exp}).  Transformer ratio (false color map) as a function of the DLW inner radius $a$ and dielectric layer thickness $b-a$ with corresponding  $|E^+|$ shown as isoclines with values quoted in MV/m for case (c) of Fig.~\ref{fig:exp}. \label{fig:perf} }
\end{figure}

We now quantify the performance of the produced current profiles to enhance beam-driven acceleration techniques by considering a drive bunch injected in a cylindrical-symmetric dielectric-lined waveguide (DLW)~\cite{gai}.  The DLW consists of a hollow dielectric cylinder with inner and outer radii $a$ and $b$. The cylinder is taken to be diamond (relative electric permittivity $\epsilon_r=5.7$); and its outer surface is contacted with a perfect conductor; see Fig.~\ref{fig:wake} (a). The measured current profiles are numerically convolved with the Green's function associated to the monopole mode to yield the axial electric field~\cite{rosinggai}. These semi-analytical calculations were  benchmarked against finite-difference time-domain electromagnetic simulations executed with {\sc vorpal}~\cite{vorpal}. The transformer ratio is numerically inferred as ${\cal R} \equiv |E_+/E_-|$ where  $E_-$ (resp. $E_+$) is the decelerating (resp. accelerating) axial electric field within (resp. behind) the electron bunch; see Fig.~\ref{fig:wake} (b).  The achieved ${\cal R}$ and $E_+$ values as the structure geometry is varied are shown in Fig.~\ref{fig:perf}. As $a \in [20,300]$~$\mu$m and $b \in a+[20,300]$~$\mu$m are varied the wavelengths of the excited wakefield modes  change. The simulations show that profiles (b) and (c) of Fig.~\ref{fig:exp} can yield values of ${\cal R} > 2$. A possible configuration with $(a,b)=(20,60)$~$\mu$m, results in  ${\cal R} \simeq 5.8$ with $E^+\simeq 0.75$~GV/m; see corresponding wake in Fig.~\ref{fig:wake} (b). Such high-field with transformer ratio significantly higher than 2 and driven by bunches produced in a superconducting linac could pave the way toward compact high-repetition-rate short-wavelength FELs~\cite{jing}. 

Finally, the proposed technique could be adapted to non-ultrarelativistic energies using a two- (or multi-) frequency version of the velocity-bunching scheme~\cite{velo}. Such an implementation would circumvent the use of a BC and would therefore be immune to CSR effects. 

In summary we proposed and experimentally demonstrated a simple method for shaping the current profile of relativistic electron bunches. The technique could be further refined by, e.g., including several harmonic frequencies. 

We are thankful to the FLASH team for the excellent technical support. We thank K. Fl\"ottmann, T. Limberg, I. Zagorodnov, E. Vogel, S. Wesch, H. Edwards, B. Faatz, K. Honkavaara, and S. Schreiber for discussions and  support. This work was sponsored by the DTRA  award HDTRA1-10-1-0051 to Northern Illinois University,  the German's Bundesministerium f\"ur Bildung und Forschung and by the DOE contract DE-AC02-07CH11359 to the Fermi research alliance LLC.

\end{document}